\documentclass[twocolumn,showpacs,prb]{revtex4}

\usepackage{graphicx}
\usepackage{dcolumn}
\usepackage{bm}

\begin{document}

\preprint{???}

\title{Molecular motions in lipid bilayers studied by the neutron backscattering technique}

\author{Maikel~C.~Rheinst\"adter}\email{rheinstaedter@ill.fr}
\affiliation{Institut Laue-Langevin, 6 rue Jules Horowitz, BP
156,38042 Grenoble Cedex 9, France}

\author{Tilo~Seydel}
\affiliation{Institut Laue-Langevin, 6 rue Jules Horowitz, BP
156,38042 Grenoble Cedex 9, France}

\author{Franz~Demmel}
\affiliation{Institut Laue-Langevin, 6 rue Jules Horowitz, BP
156,38042 Grenoble Cedex 9, France}

\author{Tim~Salditt}
\affiliation{Institut f\"{u}r R\"{o}ntgenphysik,
Georg-August-Universit\"{a}t G\"{o}ttingen, Geiststra{\ss}e 11,
37037 G\"{o}ttingen, Germany}

\date{\today}

\begin{abstract}
We report a high energy-resolution neutron backscattering study to
investigate slow motions on nanosecond time scales in highly
oriented solid supported phospholipid bilayers of the model system
DMPC -d54 (deuterated
1,2-dimyristoyl-sn-glycero-3-phoshatidylcholine), hydrated with
heavy water. This technique allows to discriminate the onset of
mobility at different length scales for the different molecular
components, as e.g.\@ the lipid acyl-chains and the hydration
water in between the membrane stacks, respectively, and provides a
benchmark test regarding the feasibility of neutron backscattering
investigations on these sample systems. We discuss freezing of the
lipid acyl-chains, as observed by this technique, and observe a
second freezing transition which we attribute to the hydration
water.
\end{abstract}


\maketitle

\section{Introduction\label{Introduction}}
%
%
\begin{figure} \centering
\resizebox{1.00\columnwidth}{!}{\rotatebox{0}{\includegraphics{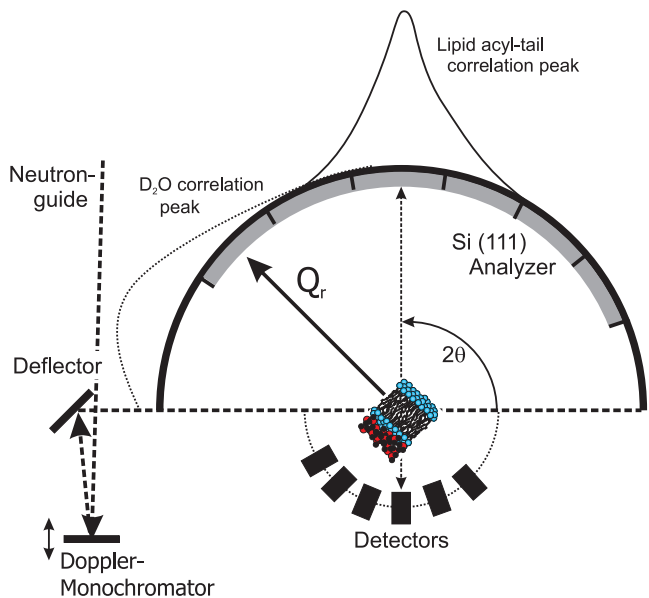}}}
\caption[]{(Color online) Schematic of the scattering geometry
(not drawn to scale). The sample was mounted with the membrane
planes vertical, i.e.\@ perpendicular to the horizontal scattering
plane of the backscattering spectrometer. The membrane planes were
oriented at well-defined angles with respect to the incident beam.
The inter acyl-chain correlation peak in the plane of the
membranes is located at $Q_r$=1.42~\AA$^{-1}$. Spatially arranged
analyzers at a sample-to-crystal distance of approx.~$1.5\,$m
allow to separately but simultaneously probe the dynamics on
different length scales. The bulk heavy water correlation peak,
which occurs at $Q_r$=2~\AA$^{-1}$, is indicated by the dotted
line.} \label{backscattering_schematic.eps}
\end{figure}
Lipid membranes as model systems for more complex biological
membranes \cite{Lipowsky:1995} cannot be understood without taking
into account the structure and dynamics of their aqueous
environment. The structure and dynamical properties of the bound
water layers next to the bilayer as well as the 'free' (bulk)
water further away from the water/lipid interface  are of
importance in understanding the thermal, elastic and transport
properties of membranes. Furthermore, the interaction between two
bilayers is mediated by the hydration water
\cite{Israelachvili:1990,Perera:1996,Katsaras:1997,Vogel:2000,Pabst:2002}.
A recent Molecular Dynamics (MD) simulation pointed out the
importance of hydration water dynamics for the understanding of
the dynamical transition of proteins \cite{Tarek:2002}.
Complementing well established structure-function relationships in
biophysics, possible dynamics-function relationships remain to
date much less elucidated. The dynamical properties of membrane
bound water present an important example in this context.
While most spectroscopic techniques such as nuclear magnetic
resonance (NMR) or dielectric spectroscopy, are limited to the
center of the Brillouin zone and probe macroscopic responses,
neutrons and within some restrictions also x-rays give unique
access to microscopic dynamics on length scales of e.g.\@
intermolecular distances. Here, we report on a high
energy-resolution neutron backscattering study to investigate slow
motions on nanosecond time scales in highly oriented solid
supported phospholipid bilayers of the model system DMPC -d54
(deuterated 1,2-dimyristoyl-sn-glycero-3-phoshatidylcholine),
hydrated with heavy water. 
The spectrum of fluctuations in biomimetic and biological
membranes covers a large range of time and length scales
\cite{Koenig:1992,Koenig:1994,Koenig:1995,Pfeiffer:1989,Pfeiffer:1993,Lindahl:2000,Lipowsky:1995,Bayerl:2000,Salditt:2000},
ranging from the long wavelength undulation and bending modes of
the bilayer with typical relaxation times of nanoseconds and
lateral length scales of several hundred lipid molecules to the
short wavelength density fluctuations in the picosecond range on
nearest neighbor distances of lipid molecules. Local dynamics in
lipid bilayers, i.e.\@ dynamics of individual lipid molecules as
vibration, rotation, libration (hindered rotation) and diffusion,
has been investigated by incoherent neutron scattering
\cite{Pfeiffer:1989,Koenig:1992,Koenig:1994} and nuclear magnetic
resonance \cite{Nevzorov:1997,Bloom:1995} to determine, e.g.\@ the
short wavelength translational and rotational diffusion constant.
Collective undulation modes have been investigated using neutron
spin-echo spectrometers \cite{Pfeiffer:1993,Takeda:1999} and
dynamical light scattering \cite{Hirn:1998}. Recently, the first
coherent inelastic scattering experiments in phospholipid bilayers
have been performed using inelastic x-ray \cite{Chen:2001} and
neutron \cite{RheinstaedterPRL:2004} scattering techniques to
determine the short wavelength dispersion relation. Note that only
scattering experiments give wave vector resolved access to
dynamical properties, what is important to associate relaxation
times with specific motions.

Information about fluctuations on mesoscopic length scales can be
inferred from x-ray lineshape analysis
\cite{Safinya:1986,Nagle:1996} in isotropic lipid dispersions.
Off-specular x-ray and neutron reflectivity from aligned phases
presents the additional advantage that the components of the
scattering vector $\vec{Q}$ can be projected onto the symmetry
axis of the membrane
\cite{Lyatskaya:2001,Salditt:2000,SaldittLangmuir:2003}. In both
examples, the time-averaged elastic scattering is studied, and
information on e.g.\@ elasticity properties and interaction forces
can be obtained.

Contrarily, dynamical properties like transport coefficients can
be inferred from direct measurements of dynamical modes. Again,
these measurements are preferably carried out in aligned phases to
preserve the unique identification of modes on the basis of the
parallel and perpendicular components $Q_r$ and $Q_z$ of the
scattering vector $\vec{Q}$, as pictured in
Fig.~\ref{backscattering_schematic.eps}. Recently
\cite{RheinstaedterPRL:2004}, we have demonstrated that collective
supra-molecular dynamics of planar lipid bilayers, notably the
dispersion relation of density modes in the lipid acyl-chains, can
be studied using the so-called three-axis neutron spectroscopy
technique giving access to an energy resolution of up to about
300~$\mu$eV . The low scattering volume of quasi two-dimensional
planar membranes and the small inelastic signal pose  particular
experimental challenges. Sample preparation and experimental
set-ups have to be adapted for inelastic experiments. In
Ref.~\onlinecite{RheinstaedterPRL:2004} the scattering volume
restriction was overcome by stacking several thousand highly
aligned bilayers. These experiments shall be complemented in the
present work by $\mu$eV energy resolved spectra, achieved by the
neutron backscattering technique. We thereby gain access to the
low energetic density fluctuations corresponding to slow motions
on nanosecond time scales. By analyzing the respective $Q$
dependence, we simultaneously probe internal length scales from 3
to 18~\AA\ and investigate freezing of the lipid acyl-chains and
of the hydration water, i.e.\@ the water layer in between the
stacked membranes. The paper is organized as follows: In the next
section, we briefly discuss hydration water in stacked planar
membranes based on literature. Sect.~\ref{Experimental} gives
details of the experimental set-ups used for neutron
backscattering and diffraction on three-axis spectrometers. The
corresponding results are presented in Sects.~\ref{Diffraction}
and \ref{Backscattering} and discussed in Sect.~\ref{Discussion}.



\section{Hydration water in multilamellar oriented membrane stacks\label{Review}}
Freezing of water in lamellar structures (DEPE, DEPC, DOPC) was
first reported by Gleeson {\em et al.} \cite{Gleeson:1994} from
x-ray diffraction. K\"onig {\em et al.} \cite{Koenig:1994} have
investigated the molecular dynamics of water in DPPC multilayers
by quasielastic neutron scattering (QENS) and nuclear magnetic
resonance (NMR). Neutron diffraction and QENS experiments on the
interaction of hydration water and purple membranes have been
reported by Lechner and Fitter and coworkers
\cite{Lechner:1998,Fitter:1999}. Additionally, Marrink {\em et
al.} have investigated the ordering of membrane water in DPPC by
Molecular Dynamics simulations (MD) \cite{Marrink:1993}.

The combination of experimental results and conclusions of the
complementary techniques and experiments and the outcome of the
simulations gives a plausible picture of the behavior of hydration
water in multilamellar oriented membrane stacks. The basic
scenario is the following \cite{Gleeson:1994}: The lamellar
spacing $d_z$ proves to be a very fine measure for the water
content in between the membrane stacks as it is sensitive to the
number of water layers (hydration shells) between two bilayers
(which have a thickness of about 3.5-4~\AA). The membrane
hydration water can be cooled well below the normal freezing point
of (bulk) water. $d_z$ shows a hysteresis in heating and cooling
cycles. When on cooling the temperature falls below the freezing
temperature of bulk water and the hydration water becomes
supercooled, $d_z$ starts to drop indicating a decreasing number
of water layers between the stacks. Ice Bragg peaks appear with an
intensity increase proportional to the decrease of $d_z$. The
authors conclude that the hydration water migrates out of the
membrane stacks when freezing and condenses as ordinary ice
outside the lamellar structure and the remaining hydration water
is in equilibrium with bulk ice. There is no crystalline ice
between lipid bilayers in an ordered multilamellar phase.

From QENS measurements at different degrees of hydration and
orientations (with the scattering vector $\vec{Q}$ parallel and
perpendicular to the plane of the bilayers), the different water
dynamics and anisotropy can be extracted \cite{Koenig:1994}. At
low hydration of the bilayers, rotation of water molecules is the
dominant motion without signs of translational diffusion. At high
hydration a translational diffusion has to be allowed to fit the
QENS spectra. The rotational degree of freedom corresponds to that
of free bulk water, while the translational diffusion appears to
be hindered, however, as compared to bulk water with a slightly
but significantly smaller diffusion constant $D$. The water
dynamics is isotropic within the length scales probed of about
10~\AA. The first hydration layers around the lipid head-groups
are tightly bound and show rotational dynamics, only. Additional
layers participate in the translational dynamics and behave like
quasifree bulk water. The authors estimate that about 40\% of the
water in a hydrated sample is tightly bound. 

The MD simulations show that the water structure is highly
perturbed by the presence of the two membrane surfaces
\cite{Marrink:1993}. The analysis of orientational polarization,
hydrogen bonding and diffusion rates of the water molecules
between the membranes reveals a strong perturbing effect, which
decays smoothly towards the middle of the water layer. Bulk-like
water is only found at a distance of 10~\AA\ away from the
interfacial plane. But even there, the hydrogen bonding pattern as
well as diffusion rates show small but significant deviations, in
agreement with the experiments.

\section{Experimental\label{Experimental}}

Deuterated DMPC -d54 (deuterated
1,2-dimyristoyl-sn-glycero-3-phoshatidylcholine) was obtained from
Avanti Polar Lipids.
Highly oriented membrane stacks were prepared by spreading a
solution of typically 25~mg/ml lipid in
trifluoroethylene/chloroform (1:1) on 2'' silicon wafers, followed
by subsequent drying in vacuum and hydration from $D_2O$ vapor
\cite{Muenster:1999}. Fifteen such wafers separated by small air
gaps were combined and aligned with respect to each other to
create a 'sandwich sample' consisting of several thousands of
highly oriented lipid bilayers (total mosaicity of about
0.6$^{\circ}$), with a total mass of about 400~mg of deuterated
DMPC.

%
%
The experiment was carried out at the cold neutron backscattering
spectrometer IN10~\cite{Alefeld:1992} at the Institut
Laue-Langevin (ILL) in its standard setup with Si(111)
monochromator and analyzer crystals corresponding to an incident
and analyzed neutron energy of 2.08~meV ($\lambda$=6.27~\AA) (as
sketched in Fig.\ref{backscattering_schematic.eps}). Neutron
backscattering was used to obtain wave vector-resolved dynamical
information with a high resolution in energy transfer of about
0.9~$\mu$eV~FWHM. The high energy resolution in neutron
backscattering results from the use of 90$^{\circ}$ Bragg angles
at both the monochromator and analyzer
crystals~\cite{MaierLeibniz:1966}. Thus in principle the precision
in the determination of the neutron wavelength is only limited by
the Darwin width of the crystals used~\cite{Darwin:1914}. In the
existing experimental implementations of the backscattering
geometry, contributions from the beam divergence and deviations
from exact 90$^{\circ}$ Bragg angles slighly deteriorate this
precision. As is indicated in
Fig.\ref{backscattering_schematic.eps}, in the IN10 spectrometer a
small deviation from exact backscattering occurs at the
monochromator (less than 1$^{\circ}$) to allow for the returning
neutrons to hit the deflector and subsequently illuminate the
sample inside the secondary spectrometer. The scattered neutrons
returning from the analyzers are detected in exact backscattering
and thus may travel twice through the sample. This does not pose a
problem since the total scattering probability by the sample is
always kept below 10\%. The incident beam is adequately pulsed,
and therefore directly scattered neutrons can be discriminated
from analyzed neutrons by their flight time.

The incident beam at the sample in our experiment was
about 3~cm wide and 4.5~cm high with a divergence corresponding to
the critical angle of Ni$^{58}$.
The sample was mounted in a hermetically sealed aluminium container
within a cryostat and hydrated from $D_2O$ vapor. Saturation of
the vapor in the voids around the lipids was assured by placing a
piece of pulp soaked in $D_2O$ within the sealed sample container.
%
%
The pulp was shielded by Cadmium to exclude any parasitic
contribution to the scattering.
The hydration was not controlled but we allowed the sample to
equilibrate for 10~h at room temperature before the measurements.
%
%
We note that the absolute level of hydration of the sample is
therefore not precisely known \footnote{A higher precision would
require a cryostated {\it in situ} absorption isotherm apparatus
which is at present not available.}, but the temperature of the
main transition agrees quite well with literature values. We argue
that the question of hydration might become more important when
discussing relaxation times and diffusion constants while in this
study we mainly concentrate on phase transitions and elastic
scattering.

Using the IN10 spectrometer, two types of measurements could be
performed: Firstly, fixed energy-window scans centered at zero
energy transfer (FEW-scans), have been recorded as a function of
the sample temperature.  This scattering intensity can be regarded
as purely elastic, within the excellent instrumental resolution
rendering motions faster than approx. 4~ns already visible.
Changes in this intensity may arise from either structural or
dynamical changes, i.e.\@ shifts of correlation peaks or freezing
of dynamical modes. In connection with diffraction data,
structural changes may be separated from dynamical effects. Thus,
from FEW-scans, information on the onset and type of molecular
mobility in the sample can be inferred and glass or melting
transitions clearly identified and assigned to corresponding
length scales by analyzing the corresponding $Q$-dependence. The
second type of measurement was performed by Doppler-shifting the
incident neutron energy through an adequate movement of the
monochromator crystal. These energy scans correspond to a time
range of motion in the sample on the order of
$10^{-9}\,s<t<\,10^{-8}\,s$.

In view of its entangled geometry, the application of the neutron
backscattering technique for probing dynamics at interfaces is
challenging. Nevertheless, we demonstrate here the feasibility of
backscattering on the lipid membrane sandwich samples. Analogously
to the three-axis experiments, the samples have been oriented in
the spectrometer to measure at wave vector transfers parallel and
perpendicular to the lipid membrane plane, respectively. The IN10
analyzers cover an angular range of approximately 20$^{\circ}$
each, resulting in a rather poor $Q$-resolution (which does not
allow to measure diffraction data to determine, e.g.\@ the
lamellar spacing $d_z$), but enhanced sensitivity for even very
small inelastic signals. Note that we are therefore not sensitive
to shifts of correlation peaks within this coarse $Q$ resolution
and corresponding structural changes of the bilayers. The six
discrete detector tubes (D1-D6) of IN10 cover a total $Q$-range of
0.34~\AA$^{-1}$ up to 1.91~\AA$^{-1}$, namely

\begin{tabular}{c|c|c}
Detector & center $Q$ value & $Q$ range\\\hline
D1 & 0.5~\AA$^{-1}$  & 0.34~\AA$^{-1}<$D1$<$0.69~\AA$^{-1}$ \\
D2 & 0.86~\AA$^{-1}$ & 0.69~\AA$^{-1}<$D2$<$1.00~\AA$^{-1}$\\
D3 & 1.18~\AA$^{-1}$ & 1.00~\AA$^{-1}<$D3$<$1.29~\AA$^{-1}$\\
D4 & 1.42~\AA$^{-1}$ & 1.29~\AA$^{-1}<$D4$<$1.54~\AA$^{-1}$\\
D5 & 1.69~\AA$^{-1}$ & 1.59~\AA$^{-1}<$D5$<$1.78~\AA$^{-1}$\\
D6 & 1.85~\AA$^{-1}$ & 1.78~\AA$^{-1}<$D6$<$1.91~\AA$^{-1}$.
\end{tabular}

The onset of molecular mobility can thus be measured separately
but simultaneously for different molecular components at different
$Q$ values and length scales. The lipid acyl-chain correlation
peak that occurs at $Q_r\simeq$1.42~\AA$^{-1}$ was mainly detected
in D4, as pictured in Fig.~\ref{backscattering_schematic.eps}.
Note that the diffraction angle of (heavy) bulk water
($Q$=2~\AA$^{-1}$) is not directly accessible on IN10 for
geometrical reasons. The detector at highest angle (D6), which is
centered at $Q$=1.85~\AA$^{-1}$, covers a $Q$ range of approx.\@
1.78~\AA$^{-1}<Q<$1.91~\AA$^{-1}$ and measures in the tails of the
broad (heavy) water correlation peak. D5 and D6 also cover the
range of the (hexagonal, $P6_3/mmc$) ice Bragg peaks (100), (002)
and (101) at $Q$ values of 1.605~\AA$^{-1}$, 1.706~\AA$^{-1}$ and
1.817~\AA$^{-1}$, respectively. We have measured all scans
(elastic and inelastic) also with the membranes rotated by
90$^{\circ}$, hence with $\vec{Q}$ perpendicular to the membrane
surface ($Q_z$), as a reference.

Complementary diffraction data have been measured on the cold and
thermal three-axis spectrometers IN12 and IN3. IN3 was equipped
with its multi analyzer detector\cite{Demmel:1998} to map out
large areas of reciprocal space simultaneously, with all analyzers
aligned to a fixed final energy of the scattered neutrons of
$E_f$=31~meV. IN12 was used with fixed $k_f$=1.75~\AA$^{-1}$
resulting in a $Q$ resolution of $\Delta Q$=0.02~\AA$^{-1}$. For
the three-axis experiments, the sample was held in a dedicated
humidity chamber and -- as for the backscattering measurements --
hydrated from the vapor phase. The use of a spectrometer to
measure diffraction (energy transfer $\Delta\omega=0$) has a
particular advantage when working with partially deuterated and/or
biological samples containing C, O, N, H (or D). As the downstream
analyzer cuts out the (within the instrumental resolution)
elastically scattered neutrons, the quasielastic contribution to
the background arising from incoherent scattering is reduced and
the signal to noise ratio drastically improved.
\subsection{Elastic neutron diffraction\label{Diffraction}}
\begin{figure} \centering
\resizebox{1.00\columnwidth}{!}{\rotatebox{0}{\includegraphics{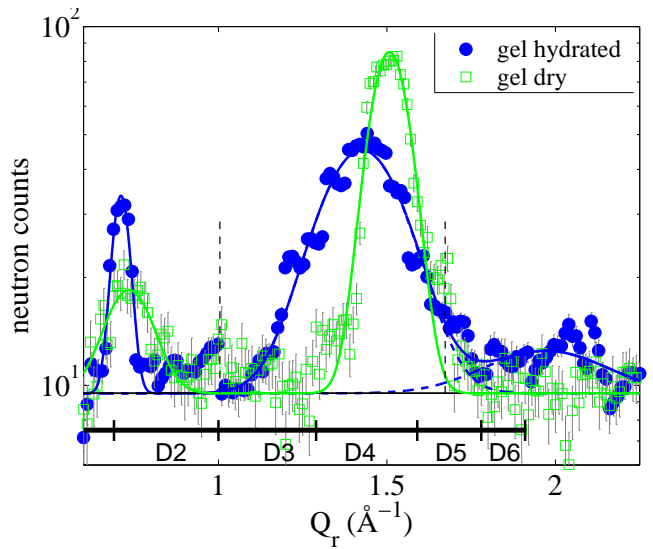}}}
\caption[]{(Color online) Diffraction data of a dry (ambient
humidity) and a hydrated sample ($T$=18$^{\circ}$C) taken on IN12
and IN3, respectively. The scans cover the $Q$ range of the
correlation peaks of the lipid acyl-chains and of water. The $Q$
coverage of the six discrete IN10 detectors D1-D6 is indicated in
the Figure. (Dotted lines mark positions of Aluminum powder
lines.)} \label{diffrac_cut.eps}
\end{figure}
\begin{figure}
\centering
\resizebox{1.00\columnwidth}{!}{\rotatebox{0}{\includegraphics{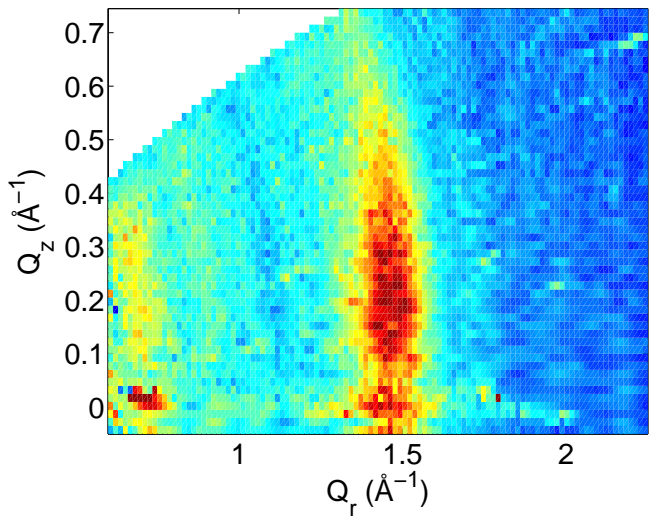}}}
\caption[]{(Color online) $Q_r/Q_z$ mapping of the diffraction
signal from the hydrated sample (gel phase, $T$=18$^{\circ}$C).
The corresponding $Q_z=0$ section is shown in
Fig.~\ref{diffrac_cut.eps} (IN3 multidetector data). The
correlation peak of the acyl-chains exhibits the shape of a
slightly bent Bragg rod, reflecting the quasi two-dimensional
liquid-like short range correlations of the acyl chain positions.}
\label{diffrac.eps}
\end{figure}
Figure~\ref{diffrac_cut.eps} shows $Q_r$ scans of DMPC, measured
on IN12 and IN3 at $T$=18$^{\circ}C$, for a dry (ambient humidity)
and a hydrated ($d_z$=58.5~\AA$^{-1}$) DMPC sample, respectively.
The measured $Q_r$ range of 0.65~\AA$^{-1}<Q_r<2.25$~\AA$^{-1}$
covers the correlation peaks of the lipid acyl-chains ($Q_r\simeq
1.42$~\AA$^{-1}$) and heavy bulk water ($Q_r\simeq 2$~\AA$^{-1}$),
which originate from the average nearest neighbor distances of
acyl-chains and water molecules, respectively. The additional peak
observed at around $Q_r\simeq 0.7$~\AA$^{-1}$ could be a
contribution of the packing of the lipid head groups in these gel
or sub-gel phases, which is usually not seen with x-rays. But the
exact assignment will need a dedicated study. As is well known,
the $Q_r$ position of the chain correlation maximum shifts to
smaller values upon hydration, reflecting the corresponding
increase of the nearest neighbor distance. An additional broad
peak appears at around $Q_r\simeq$2~\AA$^{-1}$ (FWHM
0.4~\AA$^{-1}$) and is attributed to the water layer between the
membranes. Figure~\ref{diffrac.eps} shows a mapping of the elastic
scattering intensity in the $Q_r/Q_z$ plane in the hydrated gel
phase, at $T$=18$^{\circ}$C. The intensity distribution reflects
the quasi two-dimensional liquid structure, as discussed in
Ref.~\onlinecite{Spaar:2003}. The elongated rod-like peak can be
understood by analogy with the well known crystal truncation rods
in surface crystallography. This smearing of the intensity in the
perpendicular direction is also well known from scattering of
monolayers at the air water interface. Contrarily, the weak
intensity at the water position ($Q_r=2$~\AA$^{-1}$) is truly
isotropic and distributed over a Debye-Scherrer ring.
Consequently, we expect a strong anisotropy in the backscattering
experiment at the $Q$ positions of the lipid acyl-chains, while
scattering at the water position should be isotropic.

\subsection{Backscattering \label{Backscattering}}
\begin{figure*}
\centering
\resizebox{1.00\textwidth}{!}{\rotatebox{0}{\includegraphics{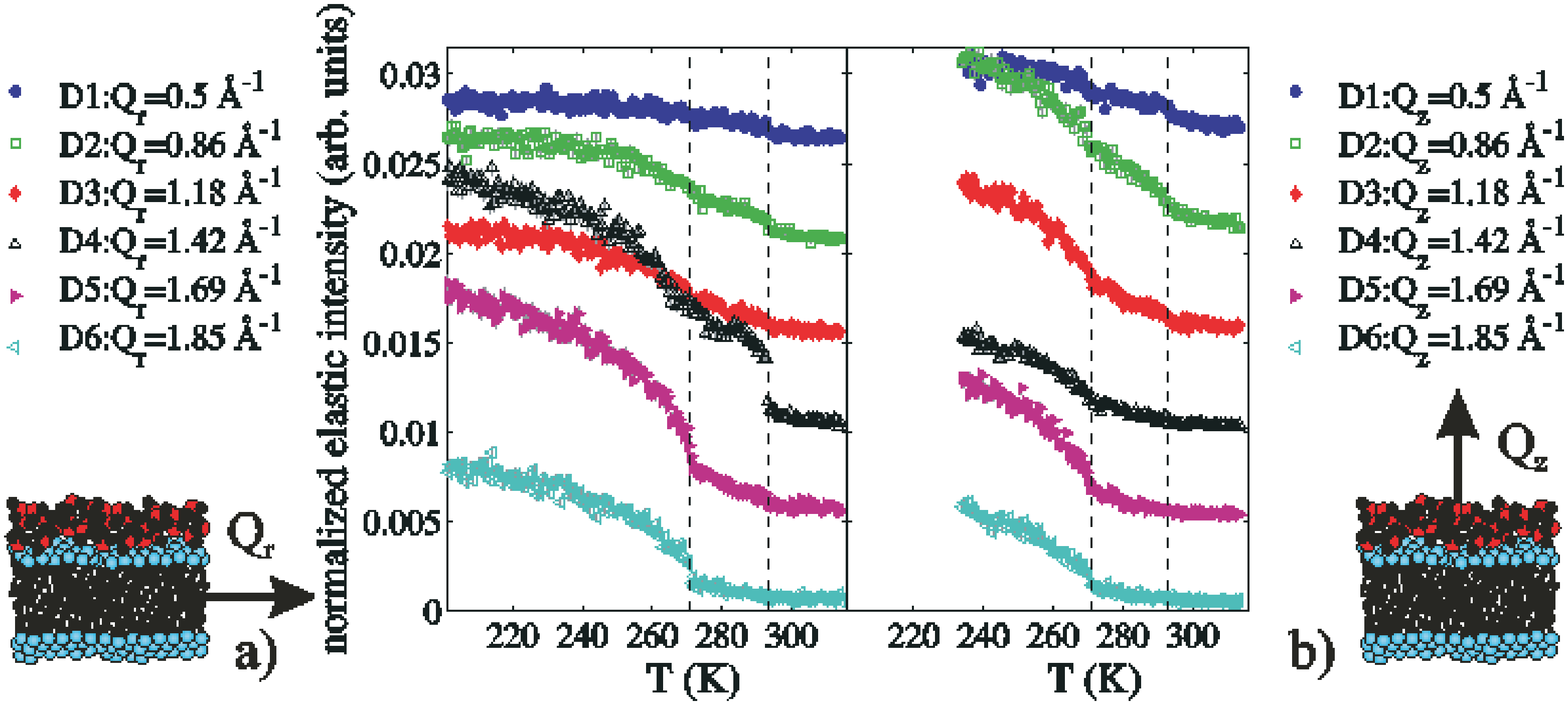}}}
\caption[]{(Color online) Fixed energy-window scans with the
scattering vector $\vec{Q}$ placed in the plane of the membranes
(a) and perpendicular to the bilayers (b). The curves are shifted
along the intensity axis for clarity. The drawings show the
orientation of the scattering vector $\vec{Q}$ with respect to the
membrane stacks. (Counting is normalized to monitor)}
\label{elastik_cdr.eps}
\end{figure*}

The intensity measured in the FEW scans is essentially equivalent
to the 'elastic structure factor' $S(Q,\omega=0)$, except for
minor temperature independent corrections, e.g.\@ due to
self-absorption or geometrical effects. In contrast to the x-ray
or neutron elastic scattering factor, this is the 'true' elastic
scattering component which is measured with the highest possible
energy resolution and minimal contribution from quasi-elastic or
inelastic scattering. Changes in this intensity are basically
sensitive to dynamical changes, since the excitation of dynamical
modes will shift scattering intensity from zero to finite energy
transfer, out of the tight fixed energy window. Note that FEW data
may include contributions from coherent and incoherent scattering,
i.e.\@ from pair- and autocorrelated scattering (collective and
local modes).

 Figure~\ref{elastik_cdr.eps}
shows FEW scans for all detectors in a temperature range of
200-315~K to map out the transition of the different molecular
components from immobile to mobile as a function of temperature
for the two set-ups, with (a) the scattering vector $\vec{Q}$
placed in the plane of the membranes and (b) perpendicular to the
bilayers. Both measurements have been taken while cooling with a
constant rate of 0.1~K/min and counting times of 5~min per point.
Note that we are therefore not sensitive to possible temperature
shifts smaller than the temperature resolution of about 0.5~K. The
data are shown as raw data without any correction, solely
normalized to the incident beam monitor. The contributions by the
Aluminum can and cryostat and Silicon wafers are in any case
temperature independent, as is the individual detector efficiency
and effects induced by the sample geometry. Note that the
wavelength used is too long to excite the Aluminum Bragg peaks.
The in-plane component ($Q_r$) in Fig.~\ref{elastik_cdr.eps} (a)
shows a distinct step in D4 at $T_{f}=$293~K, which is the
temperature of the main transition from the $P\beta'$ (gel) into
the fluid $L\alpha$ phase of the deuterated DMPC bilayers. The
transition occurs some degrees lower in the deuterated compound
(at 20.15$^{\circ}$C=293.3~K \cite{Guard:1985}) as compared to
protonated DMPC
and is detected in D4, where the maximum of the static structure
factor of the lipid-chains (the acyl-chains correlation peak)
occurs. A second, kink-like anomaly appears in detectors D5 and D6
at about $T_{fw}=$271~K. In the perpendicular direction ($Q_z$
shown in Fig.~\ref{elastik_cdr.eps} (b)), D5 and D6 display a
similar temperature behavior as for $Q_r$. Thus, the scattering at
the corresponding $Q$ position can be concluded to be isotropic.
Contrarily, the two D4 curves along $Q_z$ and along $Q_r$,
respectively, exhibit different functional forms. In particular
the step-like anomaly at $T_{f}=$293~K is lacking in the $Q_z$
curve, indicating an anisotropic scattering at this $Q$. This
strong anisotropy of the acyl-chain correlations has been
suggested by the elastic intensity distribution in
Fig.~\ref{diffrac.eps}. It is therefore not astonishing that
temperature induced changes in the dynamics (within the
experimental window of time and length scales) measured at the $Q$
position of the acyl chain peak is also anisotropic.

\begin{figure}
\centering
\resizebox{0.75\columnwidth}{!}{\rotatebox{0}{\includegraphics{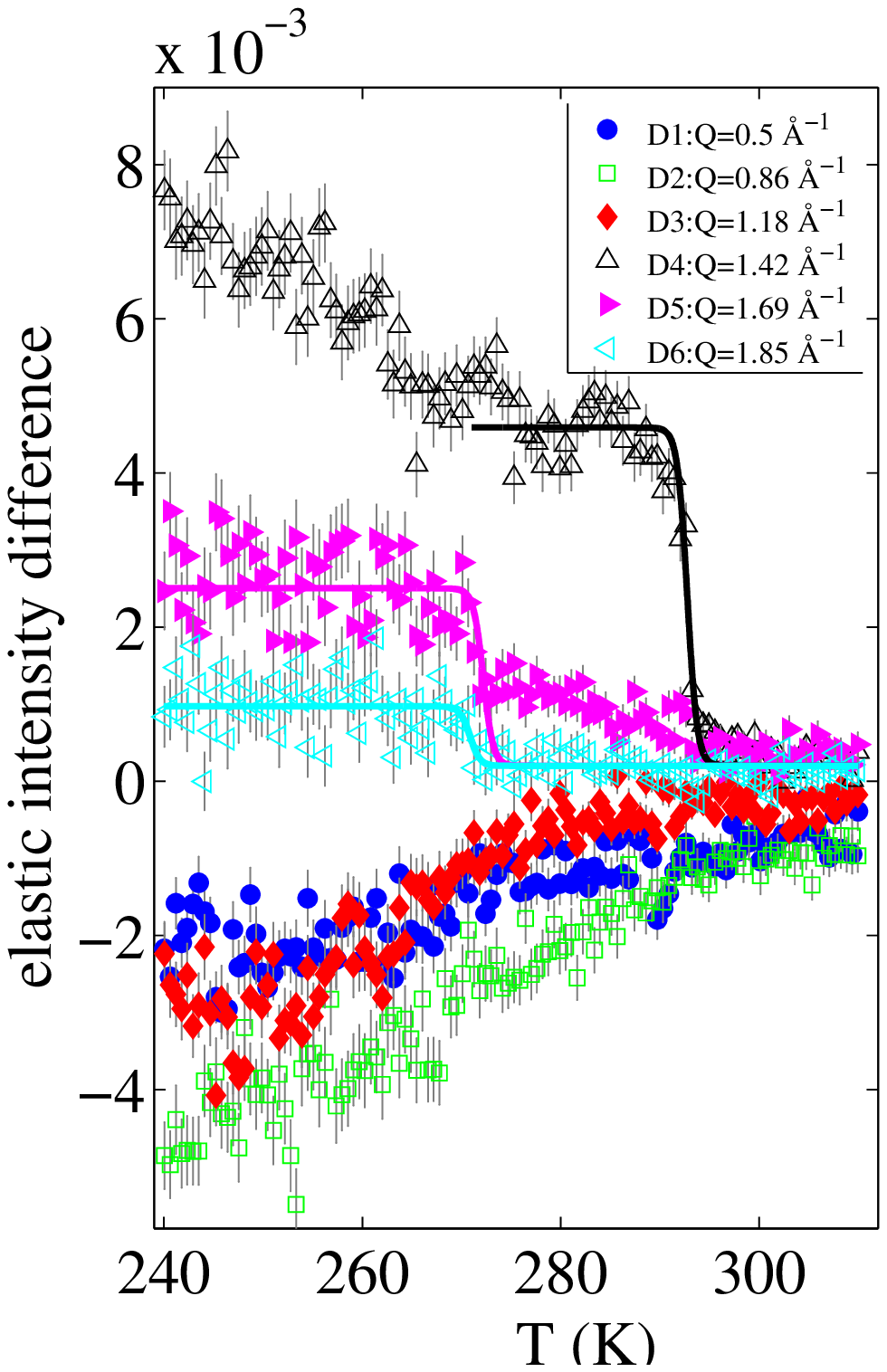}}}
\caption[]{(Color online) Anisotropy of the elastic scattering
signal (see text for explanation). The mobile-immobile transition
(within the resolution window) is clearly different for the 'lipid
chains detector' and the 'water peak detectors'. Characteristic
steps indicating the excitation of dynamical modes (or freezing of
modes upon cooling) are observed at $T_f$=293~K (attributed to
lipid acyl chains) and at $T_{fw}$=271~K (attributed to the
membrane water). Solid lines are guides to the eye. (Counting is
normalized to monitor).} \label{freezing.eps}
\end{figure}

To quantify the anisotropy of the elastic scattering, we consider
the intensity along $Q_z$ as 'background'
(Fig.~\ref{elastik_cdr.eps} (b)) and subtract these scans from the
scans in Fig.~\ref{elastik_cdr.eps} (a). Thus in
Fig.~\ref{freezing.eps} we plot the monitor normalized intensity
differences $I_{norm}(Q_r)-I_{norm}(Q_z)$. In this representation,
isotropic contributions to the signal cancel out, and anisotropic
contributions become more pronounced due to the background
subtraction. Note that this intensity difference becomes negative
if the elastic contribution along $Q_z$ is stronger than along
$Q_r$ and can no longer to be considered as a background. This is
the case for the small angle detectors D1-D3, which are expected
to pick up signal from the lamellar Bragg sheet tails
(non-specular reflectivity). As shown in Fig.~\ref{freezing.eps},
two well separated freezing steps ('immobile' within the
resolution window), of the lipid acyl-chains at $T_f=$293~K and at
about 271~K are clearly visible in detectors D4, D5 and D6.
Note that D5 obviously still shows traces of the freezing step in
the lipid acyl-chains.

\begin{figure} \centering
\resizebox{0.65\columnwidth}{!}{\rotatebox{0}{\includegraphics{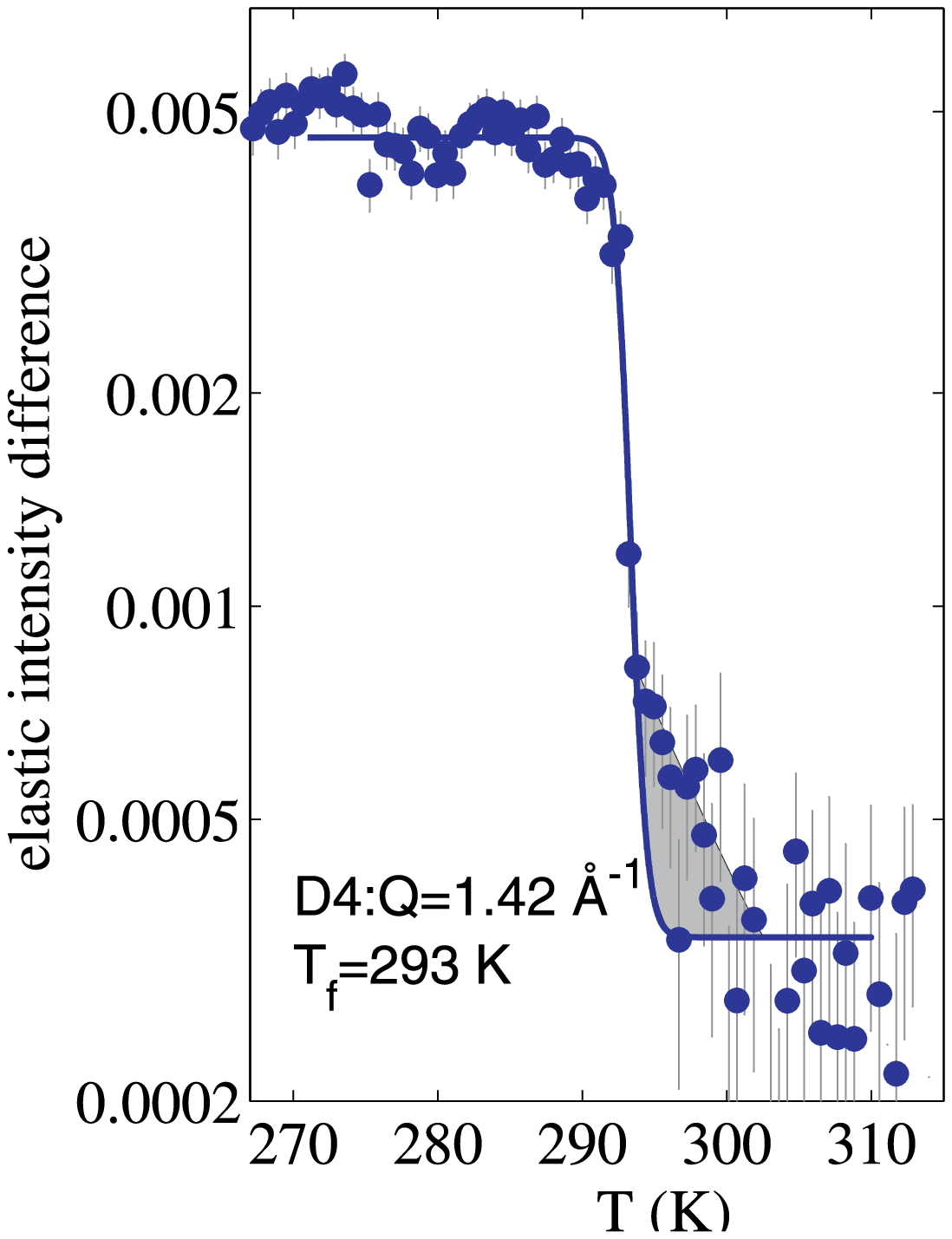}}}
\caption[]{(Color online) Freezing of dynamical modes in the lipid
acyl-chains, observed in detector D4 ($Q=1.42$~\AA$^{-1}$). There
is a sharp intensity step at $T_{f}$=293~K, corresponding to the
temperature of the gel-fluid phase transition in deuterated DMPC.
Above $T_f$ up to about 302~K there is a temperature region
characterized by  different, less steep slope (marked by the
shaded area). This temperature regime is also denoted as the
anomalous swelling regime, since the lamellar periodicity $d_z$ is
known to show a continuous swelling in this range.}
\label{freezinglipid.eps}
\end{figure}

Figure \ref{freezinglipid.eps} shows the important temperature
range around the main phase transition for the D4 detector, which
is dominated by the lipid acyl-chain correlation signal.
There is a sharp freezing step at $T_f$=293~K, corresponding to
the temperature of the gel-fluid phase transition in deuterated
DMPC. Above $T_f$ up to about 302~K, we observe a regime with a
smaller slope which coincides with the temperature range of
pseudocritical (anomalous) swelling in DMPC, as it will be
discussed below.

\begin{figure}
\centering
\resizebox{1.00\columnwidth}{!}{\rotatebox{0}{\includegraphics{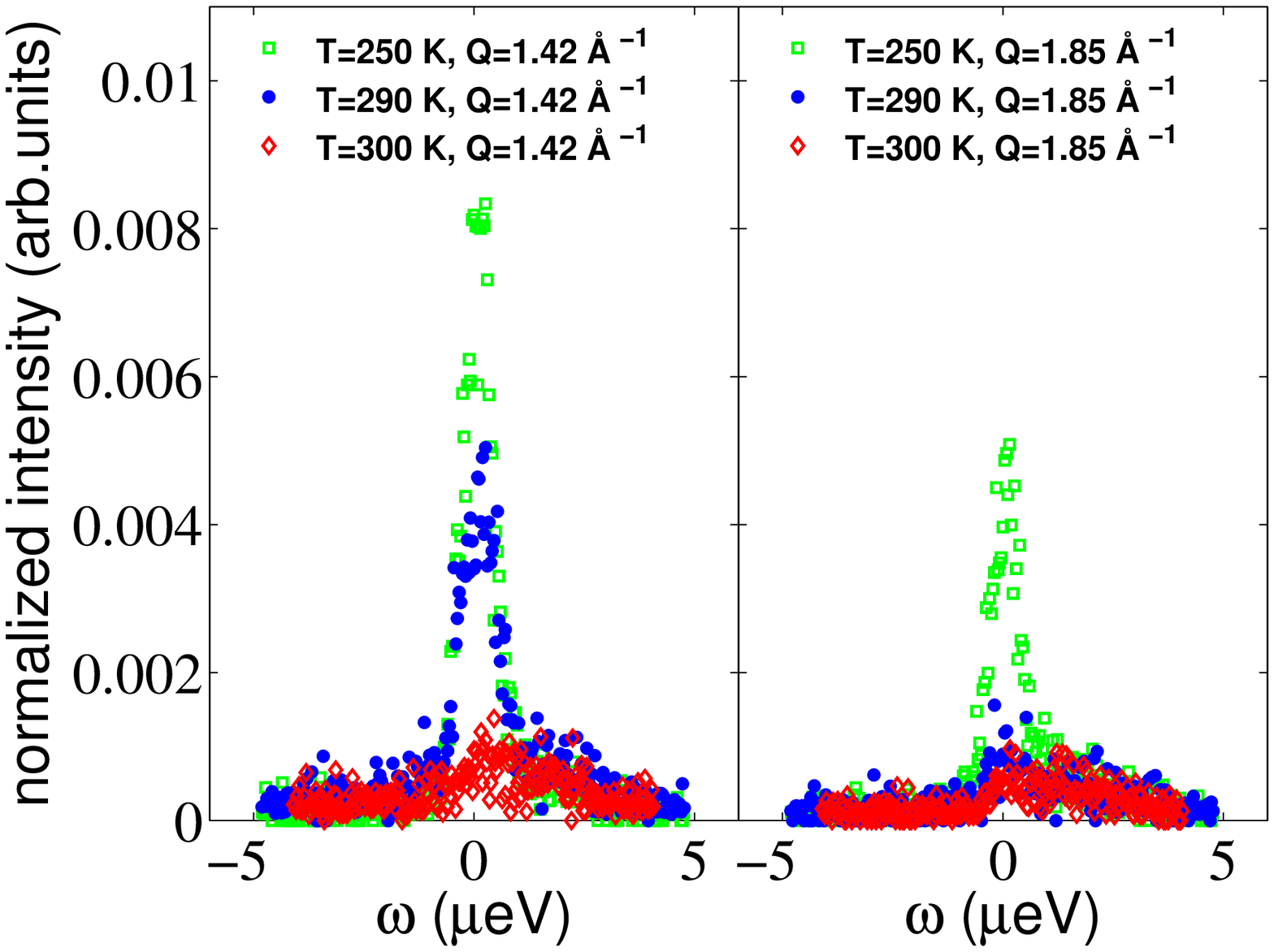}}}
\caption[]{(Color online) Energy scans at temperatures $T$=250~K,
290~K and 300~K for the $Q$-values 1.42~\AA$^{-1}$ (lipid
acyl-chain correlation peak) and $Q$=1.85~\AA$^{-1}$ (hydration
water correlation peak). At 290~K, the water is already 'mobile'
within the experimental energy resolution whereas the lipid
acyl-chains are still frozen. (Counting is normalized to
monitor)}\label{inelastik.eps}
\end{figure}
Fig.~\ref{inelastik.eps} shows representative energy transfer
scans of the deuterated DMPC membrane stack. The data have been
taken at three different temperatures, at $T$=250~K, 290~K and
300~K with a typical counting time of about 9~hours per
temperature. For these scans, the detected neutrons are sorted
into discrete channels which are evenly spaced in the measured
voltage signal arising from an induction coil at the mechanical
Doppler drive. The energy transfer is then calculated from this
voltage signal by using the known frequency of the sinosoidal
Doppler monochromator movement and the constant channel width in
terms of voltage. Therefore, we note that a systematic error may
not be completely ruled out for the energy scale of the data. From
the data, the onset of molecular mobility on the corresponding
length scales can be inferred. The presence of an elastic peak in
the spectra points to static order over the experimental time
window. Note that a fluid system does not show elastically
scattered intensity, i.e.\@ no order at infinitely long time
scales. Even within the very limited statistics, the different
dynamics are clearly visible: While the lipid acyl-chains melt
between 290 and 300~K, melting at the water position already
occurs between 250 and 290~K.

\section{Discussion\label{Discussion}}
The backscattering detectors cover the $Q$ range of the
correlation peaks of the lipid acyl-chains, bulk water and ice.
The diffraction data in Figs.~\ref{diffrac_cut.eps} and
\ref{diffrac.eps} show that the corresponding correlations lead to
signals in the neutron scattering experiments and allow to study
simultaneously lipid and water dynamics in the backscattering
experiment at the positions of the corresponding correlation
peaks. In particular, this study allows to discriminate the
immobile-to-mobile transitions of the lipid acyl-chains and the
water. The dynamics of the membrane associated water is relevant
for biological membrane function in the physiological temperature
range. The neutron backscattering technique can contribute to a
quantitative understanding of the water dynamics in membranes.
Structural and dynamical changes can lead to an increase or loss
of elastically scattered intensity in the backscattering
experiment. A structural transition or change might lead to a
shift of correlation peaks, which then move out of the respective
detector. Slowing down (freezing with respect to the instrumental
resolution and the accessible time window) of dynamical degrees of
freedom gives a quasi-static contribution and leads to an increase
in the elastically scattered intensity.
Because of the relaxed $Q$ resolution, the backscattering
experiment does not allow to probe small $Q$-shifts of the acyl
chain correlation peak \footnote{The $Q$ resolution is $Q$
dependent and increases from $\Delta Q$=0.35~\AA$^{-1}$ for D1 to
0.13~\AA$^{-1}$ for D6. At the position of the lipid correlation
peak (D4), the $Q$ resolution is $\Delta Q$=0.25~\AA$^{-1}$. We
are therefore not sensitive to the shift of the inter acyl-chain
correlation peak from the gel into the fluid phase within less
than 0.1~\AA$^{-1}$ (1.46-1.38~\AA$^{-1}$) (see e.g.\@,
Ref.~\onlinecite{RheinstaedterPRL:2004}).} and we therefore argue
that we are basically sensitive to dynamical changes.

There is a sharp freezing step in the FEW scans in the detector
located at $Q$=1.42~\AA$^{-1}$, indicating a first order
transition at $T_f$=293~K. The onset of mobility in the
acyl-chains of the lipid membranes probably deviates from a simple
melting transition, as suggested by the slow decrease of the
elastically scattered intensity with temperature in the range
250~K$<$T$<$290~K (see
Figs.~\ref{elastik_cdr.eps},\ref{freezing.eps}). The melting (or
freezing) of the lipid acyl-chains in DMPC and other lipids is
indeed of first order, but is known to show a pseudocritical
swelling, i.e.\@ a continuous change of the lamellar $d_z$-spacing
above $T_f$ (see
Refs.~\onlinecite{Nagle:1998,Zhang:1995,Chen:1997,Mason:2001}) and
also of the inter acyl-chain correlation peak
\cite{RheinstaedterPRL:2004}. The backscattering data probes the
temperature induced changes of dynamical modes with ultra high
energy but very modest $Q$ resolution. Here we observe a region
above $T_f$ of $T_f<T<T_f+9$ K, with a smaller slope in the
anisotropic elastic scattering (Fig.~\ref{freezinglipid.eps})
which coincides well with the region of pseudocritical swelling.
There is an ongoing discussion about (pseudo)criticality in lipid
bilayers and the different contributions of water layer and lipids
to the anomalous expansion. Pabst {\em et al.\@} \cite{Pabst:2003}
report from x-ray diffraction on DMPC bilayers that the anomalous
swelling is essentially the result of an expansion of the water
layer and caused by increased fluctuations. They find a softening
of the bilayer in the vicinity of $T_{f}$, i.e.\@ a (small)
increase of the Caill\'{e} fluctuation parameter $\eta_1$. The
present backscattering study allows to assign the fluctuations to
specific length scales. The critical swelling regime is thus
accompanied by significant changes in the local and collective
in-plane dynamics (diffusion, density fluctuations, undulations)
of the lipid acyl-chains in the whole range of pseudocriticality.

A quantitative analysis of the step height in
Fig.~\ref{freezinglipid.eps} shows that about 85\% of the 2D
dynamical modes in the plane of the membranes freeze at the main
transition at $T_f$. 15\% freeze already in the region of critical
swelling of the bilayers which may be accompanied by coexisting
gel and fluid microdomains, as discussed in
Ref.~\onlinecite{RheinstaedterPRL:2004} and also reported from NMR
experiments \cite{Koenig:1994}.

We attribute the second freezing transition at $T_{fw}$=271~K,
which is observed in D5 and D6, to freezing of the membrane
hydration water. Upon cooling, mobile bulk like water in the
center of the water layer may leave the layer in between the
membranes and condense as polycrystalline ice outside of the
lamellar structure. Freezing of this water leads to disappearing
of the respective dynamic modes and is responsible for the sharp
freezing step in Fig.~\ref{freezing.eps}. The polycrystalline bulk
ice then gives rise to an elastic  contribution in detectors D5
and D6, as it is observed in Fig.~\ref{elastik_cdr.eps}. Unlike
the lipid acyl-chains, the water shows a more gradual freezing.
Freezing of the hydration water is lowered by about six degrees,
as compared to pure heavy water ($^2H_2O$) at 276.97~K
\cite{Weast:1979}. The reduced number of hydrogen bonds for the
interstitial water, which was found in MD simulations
\cite{Marrink:1993}, might be responsible for the supercooling. An
interesting result of the present experiment is the fact that
after subtraction of the FEW scans along the two principal axis of
lamellar symmetry, there is a residual excess scattering
contribution also in the so-called 'water peak' detectors. In
other words, a fraction of the water exhibits anisotropic
scattering, and the freezing step of this component can be probed.

Below $T_{fw}$, all detectors in Fig.~\ref{elastik_cdr.eps} show
an increase in the elastically scattered intensity. It seems that
the membranes become 'stiffer' on all observed length scales upon
cooling. This reduction of dynamical modes may also be linked to
the decrease of the inter-bilayer distance when water freezes out.



\section{Conclusion and Outlook\label{Conclusion}}

We present the first neutron backscattering study to investigate
molecular mobility of the different molecular components in a
phospholipid membrane stack. By combining backscattering and
diffraction data, we are able to separate pure structural from
dynamical changes. The main transition from the gel to the fluid
$L\alpha$ phase is connected with freezing of collective or
diffusive motions of the acyl-chains.
In the temperature range of the so-called pseudocritical
(anomalous) swelling we find  a continuous tail in the FEW scans
at the high temperature side of the freezing step, presenting a
deviation from the simplest scenario of a first order freezing
transition. A second transition is observed at $T_{fw}\simeq$271~K
which we attribute to water molecules in the layer between the
membrane stacks.

The experiment also provides a benchmark test for inelastic
neutron studies of planar, i.e.\@ quasi two-dimensional membrane
model systems. Neutron beams for high resolution experiments are
weak even at the most intense neutron sources, and this
restriction of the low scattering intensity has been successfully
circumvented by stacking the membranes as described above. The
data are, however, still incomplete. In particular, the low count
rate in the individual detector tubes does not allow for
sufficient statistics (Fig.~\ref{inelastik.eps}). Therefore,
future experiments will be aimed at testing whether a
quasi-elastic broadening indicating slow motion on nanosecond time
scales occurs in the planar lipid membranes. In combination with
the three-axis technique, where collective short wavelength
fluctuations in the picosecond range are probed, the
backscattering technique could thus give access to {\em
relaxations} on the same length, but in the nanosecond time range.
We also note that in the present study both the lipid acyl-chains
and the water were deuterated. The experiment is therefore
predominantly sensitive to detecting collective motions arising
from coherent scattering. Further experiments will include a
selective deuteration of the acyl-chains and the membrane water,
respectively, to mask different types of mobility. In addition,
structural information will be obtained by using diffraction
detectors in parallel with the energy-discriminating
backscattering detectors. The position of the reflectivity Bragg
peaks and the lipid acyl-chain correlation peak will allow
simultaneous detection of  the lamellar spacing $d_z$ and
thickness of the water layer $d_w$, as well as the ordering of
lipids in the plane of the membranes.

{\bf Acknowledgement:} We are grateful to C.~Ollinger and
T.~Gronemann (Institut f\"ur R\"ontgenphysik, G\"ottingen) for
help with the sample preparation and to M.~Elender (ILL) for
technical and engineering support.

\bibliography{./dmpc_BackscatteringPRE}

\begin{thebibliography}{42}
\expandafter\ifx\csname natexlab\endcsname\relax\def\natexlab#1{#1}\fi
\expandafter\ifx\csname bibnamefont\endcsname\relax
  \def\bibnamefont#1{#1}\fi
\expandafter\ifx\csname bibfnamefont\endcsname\relax
  \def\bibfnamefont#1{#1}\fi
\expandafter\ifx\csname citenamefont\endcsname\relax
  \def\citenamefont#1{#1}\fi
\expandafter\ifx\csname url\endcsname\relax
  \def\url#1{\texttt{#1}}\fi
\expandafter\ifx\csname urlprefix\endcsname\relax\def\urlprefix{URL }\fi
\providecommand{\bibinfo}[2]{#2}
\providecommand{\eprint}[2][]{\url{#2}}

\bibitem[{\citenamefont{Lipowsky and Sackmann}(1995)}]{Lipowsky:1995}
\bibinfo{editor}{\bibfnamefont{R.}~\bibnamefont{Lipowsky}} \bibnamefont{and}
  \bibinfo{editor}{\bibfnamefont{E.}~\bibnamefont{Sackmann}}, eds.,
  \emph{\bibinfo{title}{Structure and Dynamics of Membranes}},
  vol.~\bibinfo{volume}{1} of \emph{\bibinfo{series}{Handbook of Biological
  Physics}} (\bibinfo{publisher}{Elsevier, North-Holland},
  \bibinfo{address}{Amsterdam}, \bibinfo{year}{1995}).

\bibitem[{\citenamefont{Israelachvili and
  Wennerstroem}(1990)}]{Israelachvili:1990}
\bibinfo{author}{\bibfnamefont{J.~N.} \bibnamefont{Israelachvili}}
  \bibnamefont{and}
  \bibinfo{author}{\bibfnamefont{H.}~\bibnamefont{Wennerstroem}},
  \bibinfo{journal}{Langmuir} \textbf{\bibinfo{volume}{6}},
  \bibinfo{pages}{873} (\bibinfo{year}{1990}).

\bibitem[{\citenamefont{Perera et~al.}(1996)\citenamefont{Perera, Essmann, and
  Berkowitz}}]{Perera:1996}
\bibinfo{author}{\bibfnamefont{L.}~\bibnamefont{Perera}},
  \bibinfo{author}{\bibfnamefont{U.}~\bibnamefont{Essmann}}, \bibnamefont{and}
  \bibinfo{author}{\bibfnamefont{M.}~\bibnamefont{Berkowitz}},
  \bibinfo{journal}{Langmuir} \textbf{\bibinfo{volume}{12}},
  \bibinfo{pages}{2625} (\bibinfo{year}{1996}).

\bibitem[{\citenamefont{Katsaras and Jeffrey}(1997)}]{Katsaras:1997}
\bibinfo{author}{\bibfnamefont{J.}~\bibnamefont{Katsaras}} \bibnamefont{and}
  \bibinfo{author}{\bibfnamefont{K.~R.} \bibnamefont{Jeffrey}},
  \bibinfo{journal}{Europhys. Lett.} \textbf{\bibinfo{volume}{38}},
  \bibinfo{pages}{43} (\bibinfo{year}{1997}).

\bibitem[{\citenamefont{Vogel et~al.}(2000)\citenamefont{Vogel, M\"unster,
  Fenzl, and Salditt}}]{Vogel:2000}
\bibinfo{author}{\bibfnamefont{M.}~\bibnamefont{Vogel}},
  \bibinfo{author}{\bibfnamefont{C.}~\bibnamefont{M\"unster}},
  \bibinfo{author}{\bibfnamefont{W.}~\bibnamefont{Fenzl}}, \bibnamefont{and}
  \bibinfo{author}{\bibfnamefont{T.}~\bibnamefont{Salditt}},
  \bibinfo{journal}{Phys. Rev. Lett.} \textbf{\bibinfo{volume}{84}},
  \bibinfo{pages}{390} (\bibinfo{year}{2000}).

\bibitem[{\citenamefont{Pabst et~al.}(2002)\citenamefont{Pabst, Katsaras, and
  Raghunathan}}]{Pabst:2002}
\bibinfo{author}{\bibfnamefont{G.}~\bibnamefont{Pabst}},
  \bibinfo{author}{\bibfnamefont{J.}~\bibnamefont{Katsaras}}, \bibnamefont{and}
  \bibinfo{author}{\bibfnamefont{V.~A.} \bibnamefont{Raghunathan}},
  \bibinfo{journal}{Phys. Rev. Lett.} \textbf{\bibinfo{volume}{88}},
  \bibinfo{pages}{128101} (\bibinfo{year}{2002}).

\bibitem[{\citenamefont{Tarek and Tobias}(2002)}]{Tarek:2002}
\bibinfo{author}{\bibfnamefont{M.}~\bibnamefont{Tarek}} \bibnamefont{and}
  \bibinfo{author}{\bibfnamefont{D.}~\bibnamefont{Tobias}},
  \bibinfo{journal}{Phys. Rev. Lett.} \textbf{\bibinfo{volume}{88}},
  \bibinfo{pages}{138101} (\bibinfo{year}{2002}).

\bibitem[{\citenamefont{Salditt}(2000)}]{Salditt:2000}
\bibinfo{author}{\bibfnamefont{T.}~\bibnamefont{Salditt}},
  \bibinfo{journal}{Curr Opin Colloid In} \textbf{\bibinfo{volume}{5}},
  \bibinfo{pages}{19} (\bibinfo{year}{2000}).

\bibitem[{\citenamefont{K\"onig et~al.}(1994)\citenamefont{K\"onig, Sackmann,
  Richter, Zorn, Carlile, and Bayerls}}]{Koenig:1994}
\bibinfo{author}{\bibfnamefont{S.}~\bibnamefont{K\"onig}},
  \bibinfo{author}{\bibfnamefont{E.}~\bibnamefont{Sackmann}},
  \bibinfo{author}{\bibfnamefont{D.}~\bibnamefont{Richter}},
  \bibinfo{author}{\bibfnamefont{R.}~\bibnamefont{Zorn}},
  \bibinfo{author}{\bibfnamefont{C.}~\bibnamefont{Carlile}}, \bibnamefont{and}
  \bibinfo{author}{\bibfnamefont{T.}~\bibnamefont{Bayerls}},
  \bibinfo{journal}{J. Chem. Phys.} \textbf{\bibinfo{volume}{100}},
  \bibinfo{pages}{3307} (\bibinfo{year}{1994}).

\bibitem[{\citenamefont{Pfeiffer et~al.}(1989)\citenamefont{Pfeiffer, Henkel,
  Sackmann, and Knorr}}]{Pfeiffer:1989}
\bibinfo{author}{\bibfnamefont{W.}~\bibnamefont{Pfeiffer}},
  \bibinfo{author}{\bibfnamefont{T.}~\bibnamefont{Henkel}},
  \bibinfo{author}{\bibfnamefont{E.}~\bibnamefont{Sackmann}}, \bibnamefont{and}
  \bibinfo{author}{\bibfnamefont{W.}~\bibnamefont{Knorr}},
  \bibinfo{journal}{Europhys. Lett.} \textbf{\bibinfo{volume}{8}},
  \bibinfo{pages}{201} (\bibinfo{year}{1989}).

\bibitem[{\citenamefont{Lindahl and Edholm}(2000)}]{Lindahl:2000}
\bibinfo{author}{\bibfnamefont{E.}~\bibnamefont{Lindahl}} \bibnamefont{and}
  \bibinfo{author}{\bibfnamefont{O.}~\bibnamefont{Edholm}},
  \bibinfo{journal}{Biophys. J.} \textbf{\bibinfo{volume}{79}},
  \bibinfo{pages}{426} (\bibinfo{year}{2000}).

\bibitem[{\citenamefont{Bayerl}(2000)}]{Bayerl:2000}
\bibinfo{author}{\bibfnamefont{T.}~\bibnamefont{Bayerl}},
  \bibinfo{journal}{Curr Opin Colloid In} \textbf{\bibinfo{volume}{5}},
  \bibinfo{pages}{232} (\bibinfo{year}{2000}).

\bibitem[{\citenamefont{K\"onig et~al.}(1992)\citenamefont{K\"onig, Pfeiffer,
  Bayerl, Richter, and Sackmann}}]{Koenig:1992}
\bibinfo{author}{\bibfnamefont{S.}~\bibnamefont{K\"onig}},
  \bibinfo{author}{\bibfnamefont{W.}~\bibnamefont{Pfeiffer}},
  \bibinfo{author}{\bibfnamefont{T.}~\bibnamefont{Bayerl}},
  \bibinfo{author}{\bibfnamefont{D.}~\bibnamefont{Richter}}, \bibnamefont{and}
  \bibinfo{author}{\bibfnamefont{E.}~\bibnamefont{Sackmann}},
  \bibinfo{journal}{J. Phys. II France} \textbf{\bibinfo{volume}{2}},
  \bibinfo{pages}{1589} (\bibinfo{year}{1992}).

\bibitem[{\citenamefont{K\"onig et~al.}(1995)\citenamefont{K\"onig, Bayerl,
  Coddens, Richter, and Sackmann}}]{Koenig:1995}
\bibinfo{author}{\bibfnamefont{S.}~\bibnamefont{K\"onig}},
  \bibinfo{author}{\bibfnamefont{T.}~\bibnamefont{Bayerl}},
  \bibinfo{author}{\bibfnamefont{G.}~\bibnamefont{Coddens}},
  \bibinfo{author}{\bibfnamefont{D.}~\bibnamefont{Richter}}, \bibnamefont{and}
  \bibinfo{author}{\bibfnamefont{E.}~\bibnamefont{Sackmann}},
  \bibinfo{journal}{Biophys. J.} \textbf{\bibinfo{volume}{68}},
  \bibinfo{pages}{1871} (\bibinfo{year}{1995}).

\bibitem[{\citenamefont{Pfeiffer et~al.}(1993)\citenamefont{Pfeiffer, K\"onig,
  Legrand, Bayerl, Richter, and Sackmann}}]{Pfeiffer:1993}
\bibinfo{author}{\bibfnamefont{W.}~\bibnamefont{Pfeiffer}},
  \bibinfo{author}{\bibfnamefont{S.}~\bibnamefont{K\"onig}},
  \bibinfo{author}{\bibfnamefont{J.}~\bibnamefont{Legrand}},
  \bibinfo{author}{\bibfnamefont{T.}~\bibnamefont{Bayerl}},
  \bibinfo{author}{\bibfnamefont{D.}~\bibnamefont{Richter}}, \bibnamefont{and}
  \bibinfo{author}{\bibfnamefont{E.}~\bibnamefont{Sackmann}},
  \bibinfo{journal}{Europhys. Lett.} \textbf{\bibinfo{volume}{23}},
  \bibinfo{pages}{457} (\bibinfo{year}{1993}).

\bibitem[{\citenamefont{Nevzorov and Brown}(1997)}]{Nevzorov:1997}
\bibinfo{author}{\bibfnamefont{A.}~\bibnamefont{Nevzorov}} \bibnamefont{and}
  \bibinfo{author}{\bibfnamefont{M.}~\bibnamefont{Brown}}, \bibinfo{journal}{J.
  Chem. Phys.} \textbf{\bibinfo{volume}{107}}, \bibinfo{pages}{10288}
  (\bibinfo{year}{1997}).

\bibitem[{\citenamefont{Bloom and Bayerl}(1995)}]{Bloom:1995}
\bibinfo{author}{\bibfnamefont{M.}~\bibnamefont{Bloom}} \bibnamefont{and}
  \bibinfo{author}{\bibfnamefont{T.}~\bibnamefont{Bayerl}},
  \bibinfo{journal}{Can. J. Phys.} \textbf{\bibinfo{volume}{73}},
  \bibinfo{pages}{687} (\bibinfo{year}{1995}).

\bibitem[{\citenamefont{Takeda et~al.}(1999)\citenamefont{Takeda, Kawabata,
  Seto, Komura, Gosh, Nagao, and Okuhara}}]{Takeda:1999}
\bibinfo{author}{\bibfnamefont{T.}~\bibnamefont{Takeda}},
  \bibinfo{author}{\bibfnamefont{Y.}~\bibnamefont{Kawabata}},
  \bibinfo{author}{\bibfnamefont{H.}~\bibnamefont{Seto}},
  \bibinfo{author}{\bibfnamefont{S.}~\bibnamefont{Komura}},
  \bibinfo{author}{\bibfnamefont{S.}~\bibnamefont{Gosh}},
  \bibinfo{author}{\bibfnamefont{M.}~\bibnamefont{Nagao}}, \bibnamefont{and}
  \bibinfo{author}{\bibfnamefont{D.}~\bibnamefont{Okuhara}},
  \bibinfo{journal}{J. Phys. Chem. Solids} \textbf{\bibinfo{volume}{60}},
  \bibinfo{pages}{1375} (\bibinfo{year}{1999}).

\bibitem[{\citenamefont{Hirn et~al.}(1998)\citenamefont{Hirn, Bayerl, R\"adler,
  and Sackmann}}]{Hirn:1998}
\bibinfo{author}{\bibfnamefont{R.}~\bibnamefont{Hirn}},
  \bibinfo{author}{\bibfnamefont{T.}~\bibnamefont{Bayerl}},
  \bibinfo{author}{\bibfnamefont{J.}~\bibnamefont{R\"adler}}, \bibnamefont{and}
  \bibinfo{author}{\bibfnamefont{E.}~\bibnamefont{Sackmann}},
  \bibinfo{journal}{Faraday Discuss.} \textbf{\bibinfo{volume}{111}},
  \bibinfo{pages}{17} (\bibinfo{year}{1998}).

\bibitem[{\citenamefont{Chen et~al.}(2001)\citenamefont{Chen, Liao, Huang,
  Weiss, Bellisent-Funel, and Sette}}]{Chen:2001}
\bibinfo{author}{\bibfnamefont{S.}~\bibnamefont{Chen}},
  \bibinfo{author}{\bibfnamefont{C.}~\bibnamefont{Liao}},
  \bibinfo{author}{\bibfnamefont{H.}~\bibnamefont{Huang}},
  \bibinfo{author}{\bibfnamefont{T.}~\bibnamefont{Weiss}},
  \bibinfo{author}{\bibfnamefont{M.}~\bibnamefont{Bellisent-Funel}},
  \bibnamefont{and} \bibinfo{author}{\bibfnamefont{F.}~\bibnamefont{Sette}},
  \bibinfo{journal}{Phys. Rev. Lett.} \textbf{\bibinfo{volume}{86}},
  \bibinfo{pages}{740} (\bibinfo{year}{2001}).

\bibitem[{\citenamefont{Rheinst\"adter
  et~al.}(2004)\citenamefont{Rheinst\"adter, Ollinger, Fragneto, Demmel, and
  Salditt}}]{RheinstaedterPRL:2004}
\bibinfo{author}{\bibfnamefont{M.}~\bibnamefont{Rheinst\"adter}},
  \bibinfo{author}{\bibfnamefont{C.}~\bibnamefont{Ollinger}},
  \bibinfo{author}{\bibfnamefont{G.}~\bibnamefont{Fragneto}},
  \bibinfo{author}{\bibfnamefont{F.}~\bibnamefont{Demmel}}, \bibnamefont{and}
  \bibinfo{author}{\bibfnamefont{T.}~\bibnamefont{Salditt}},
  \bibinfo{journal}{Phys. Rev. Lett.} \textbf{\bibinfo{volume}{93}},
  \bibinfo{pages}{108107} (\bibinfo{year}{2004}).

\bibitem[{\citenamefont{Safinya et~al.}(1986)\citenamefont{Safinya, Roux,
  Smith, Sinha, Dimon, Clark, and Bellocq}}]{Safinya:1986}
\bibinfo{author}{\bibfnamefont{C.~R.} \bibnamefont{Safinya}},
  \bibinfo{author}{\bibfnamefont{D.}~\bibnamefont{Roux}},
  \bibinfo{author}{\bibfnamefont{G.~S.} \bibnamefont{Smith}},
  \bibinfo{author}{\bibfnamefont{S.~K.} \bibnamefont{Sinha}},
  \bibinfo{author}{\bibfnamefont{P.}~\bibnamefont{Dimon}},
  \bibinfo{author}{\bibfnamefont{N.~A.} \bibnamefont{Clark}}, \bibnamefont{and}
  \bibinfo{author}{\bibfnamefont{A.~M.} \bibnamefont{Bellocq}},
  \bibinfo{journal}{Phys. Rev. Lett.} \textbf{\bibinfo{volume}{57}},
  \bibinfo{pages}{2718} (\bibinfo{year}{1986}).

\bibitem[{\citenamefont{Nagle et~al.}(1996)\citenamefont{Nagle, Zhang,
  Tristram-Nagle, Sun, Petrache, and Suter}}]{Nagle:1996}
\bibinfo{author}{\bibfnamefont{J.}~\bibnamefont{Nagle}},
  \bibinfo{author}{\bibfnamefont{R.}~\bibnamefont{Zhang}},
  \bibinfo{author}{\bibfnamefont{S.}~\bibnamefont{Tristram-Nagle}},
  \bibinfo{author}{\bibfnamefont{W.}~\bibnamefont{Sun}},
  \bibinfo{author}{\bibfnamefont{H.}~\bibnamefont{Petrache}}, \bibnamefont{and}
  \bibinfo{author}{\bibfnamefont{R.}~\bibnamefont{Suter}},
  \bibinfo{journal}{Biophys. J.} \textbf{\bibinfo{volume}{70}},
  \bibinfo{pages}{1419} (\bibinfo{year}{1996}).

\bibitem[{\citenamefont{Lyatskaya et~al.}(2001)\citenamefont{Lyatskaya, Liu,
  Tristram-Nagle, Katsaras, and Nagle}}]{Lyatskaya:2001}
\bibinfo{author}{\bibfnamefont{Y.}~\bibnamefont{Lyatskaya}},
  \bibinfo{author}{\bibfnamefont{Y.}~\bibnamefont{Liu}},
  \bibinfo{author}{\bibfnamefont{S.}~\bibnamefont{Tristram-Nagle}},
  \bibinfo{author}{\bibfnamefont{J.}~\bibnamefont{Katsaras}}, \bibnamefont{and}
  \bibinfo{author}{\bibfnamefont{J.~F.} \bibnamefont{Nagle}},
  \bibinfo{journal}{Phys. Rev. E} \textbf{\bibinfo{volume}{63}},
  \bibinfo{pages}{011907} (\bibinfo{year}{2001}).

\bibitem[{\citenamefont{Salditt et~al.}(2003)\citenamefont{Salditt, M\"unster,
  Mennicke, Ollinger, and Fragneto}}]{SaldittLangmuir:2003}
\bibinfo{author}{\bibfnamefont{T.}~\bibnamefont{Salditt}},
  \bibinfo{author}{\bibfnamefont{C.}~\bibnamefont{M\"unster}},
  \bibinfo{author}{\bibfnamefont{U.}~\bibnamefont{Mennicke}},
  \bibinfo{author}{\bibfnamefont{C.}~\bibnamefont{Ollinger}}, \bibnamefont{and}
  \bibinfo{author}{\bibfnamefont{G.}~\bibnamefont{Fragneto}},
  \bibinfo{journal}{Langmuir} \textbf{\bibinfo{volume}{19}},
  \bibinfo{pages}{7703} (\bibinfo{year}{2003}).

\bibitem[{\citenamefont{Gleeson et~al.}(1994)\citenamefont{Gleeson, Erramilli,
  and Gruner}}]{Gleeson:1994}
\bibinfo{author}{\bibfnamefont{J.}~\bibnamefont{Gleeson}},
  \bibinfo{author}{\bibfnamefont{S.}~\bibnamefont{Erramilli}},
  \bibnamefont{and} \bibinfo{author}{\bibfnamefont{S.}~\bibnamefont{Gruner}},
  \bibinfo{journal}{Biophys. J.} \textbf{\bibinfo{volume}{67}},
  \bibinfo{pages}{706} (\bibinfo{year}{1994}).

\bibitem[{\citenamefont{Lechner et~al.}(1998)\citenamefont{Lechner, Fitter,
  Dencher, and Hau{\ss}}}]{Lechner:1998}
\bibinfo{author}{\bibfnamefont{R.}~\bibnamefont{Lechner}},
  \bibinfo{author}{\bibfnamefont{J.}~\bibnamefont{Fitter}},
  \bibinfo{author}{\bibfnamefont{N.}~\bibnamefont{Dencher}}, \bibnamefont{and}
  \bibinfo{author}{\bibfnamefont{T.}~\bibnamefont{Hau{\ss}}},
  \bibinfo{journal}{J. Mol. Biol.} \textbf{\bibinfo{volume}{277}},
  \bibinfo{pages}{593} (\bibinfo{year}{1998}).

\bibitem[{\citenamefont{Fitter et~al.}(1999)\citenamefont{Fitter, Lechner, and
  Dencher}}]{Fitter:1999}
\bibinfo{author}{\bibfnamefont{J.}~\bibnamefont{Fitter}},
  \bibinfo{author}{\bibfnamefont{R.}~\bibnamefont{Lechner}}, \bibnamefont{and}
  \bibinfo{author}{\bibfnamefont{N.}~\bibnamefont{Dencher}},
  \bibinfo{journal}{J. Phys. Chem.} \textbf{\bibinfo{volume}{103}},
  \bibinfo{pages}{8036} (\bibinfo{year}{1999}).

\bibitem[{\citenamefont{Marrink et~al.}(1993)\citenamefont{Marrink, Berkowitz,
  and Berendsen}}]{Marrink:1993}
\bibinfo{author}{\bibfnamefont{S.-J.} \bibnamefont{Marrink}},
  \bibinfo{author}{\bibfnamefont{M.}~\bibnamefont{Berkowitz}},
  \bibnamefont{and}
  \bibinfo{author}{\bibfnamefont{H.}~\bibnamefont{Berendsen}},
  \bibinfo{journal}{Langmuir} \textbf{\bibinfo{volume}{9}},
  \bibinfo{pages}{3122} (\bibinfo{year}{1993}).

\bibitem[{\citenamefont{M\"unster et~al.}(1999)\citenamefont{M\"unster,
  Salditt, Vogel, Siebrecht, and Peisl}}]{Muenster:1999}
\bibinfo{author}{\bibfnamefont{C.}~\bibnamefont{M\"unster}},
  \bibinfo{author}{\bibfnamefont{T.}~\bibnamefont{Salditt}},
  \bibinfo{author}{\bibfnamefont{M.}~\bibnamefont{Vogel}},
  \bibinfo{author}{\bibfnamefont{R.}~\bibnamefont{Siebrecht}},
  \bibnamefont{and} \bibinfo{author}{\bibfnamefont{J.}~\bibnamefont{Peisl}},
  \bibinfo{journal}{Europhys. Lett.} \textbf{\bibinfo{volume}{46}},
  \bibinfo{pages}{486} (\bibinfo{year}{1999}).

\bibitem[{\citenamefont{Alefeld et~al.}(1992)\citenamefont{Alefeld, Springer,
  and Heidemann}}]{Alefeld:1992}
\bibinfo{author}{\bibfnamefont{B.}~\bibnamefont{Alefeld}},
  \bibinfo{author}{\bibfnamefont{T.}~\bibnamefont{Springer}}, \bibnamefont{and}
  \bibinfo{author}{\bibfnamefont{A.}~\bibnamefont{Heidemann}},
  \bibinfo{journal}{Nucl. Sci. Eng.} \textbf{\bibinfo{volume}{110}},
  \bibinfo{pages}{84} (\bibinfo{year}{1992}).

\bibitem[{\citenamefont{Maier-Leibniz}(1966)}]{MaierLeibniz:1966}
\bibinfo{author}{\bibfnamefont{H.}~\bibnamefont{Maier-Leibniz}},
  \bibinfo{journal}{Nukleonik} \textbf{\bibinfo{volume}{8}},
  \bibinfo{pages}{61} (\bibinfo{year}{1966}).

\bibitem[{\citenamefont{Darwin}(1914)}]{Darwin:1914}
\bibinfo{author}{\bibfnamefont{C.}~\bibnamefont{Darwin}},
  \bibinfo{journal}{Philos. Mag.} \textbf{\bibinfo{volume}{27}},
  \bibinfo{pages}{315 675} (\bibinfo{year}{1914}).

\bibitem[{\citenamefont{Demmel et~al.}(1998)\citenamefont{Demmel, Fleischmann,
  and Gl\"aser}}]{Demmel:1998}
\bibinfo{author}{\bibfnamefont{F.}~\bibnamefont{Demmel}},
  \bibinfo{author}{\bibfnamefont{A.}~\bibnamefont{Fleischmann}},
  \bibnamefont{and} \bibinfo{author}{\bibfnamefont{W.}~\bibnamefont{Gl\"aser}},
  \bibinfo{journal}{Nucl. Instrum. Methods Phys. Res., Sect. A}
  \textbf{\bibinfo{volume}{416}}, \bibinfo{pages}{115} (\bibinfo{year}{1998}).

\bibitem[{\citenamefont{Spaar and Salditt}(2003)}]{Spaar:2003}
\bibinfo{author}{\bibfnamefont{A.}~\bibnamefont{Spaar}} \bibnamefont{and}
  \bibinfo{author}{\bibfnamefont{T.}~\bibnamefont{Salditt}},
  \bibinfo{journal}{Biophys. J.} \textbf{\bibinfo{volume}{85}},
  \bibinfo{pages}{1576} (\bibinfo{year}{2003}).

\bibitem[{\citenamefont{Guard-Friar et~al.}(1985)\citenamefont{Guard-Friar,
  Chen, and Engle}}]{Guard:1985}
\bibinfo{author}{\bibfnamefont{D.}~\bibnamefont{Guard-Friar}},
  \bibinfo{author}{\bibfnamefont{C.-H.} \bibnamefont{Chen}}, \bibnamefont{and}
  \bibinfo{author}{\bibfnamefont{A.}~\bibnamefont{Engle}}, \bibinfo{journal}{J.
  Phys. Chem.} \textbf{\bibinfo{volume}{89}}, \bibinfo{pages}{1810}
  (\bibinfo{year}{1985}).

\bibitem[{\citenamefont{Nagle et~al.}(1998)\citenamefont{Nagle, Petrache,
  Gouliaev, Tristram-Nagle, Liu, Suter, and Gawrisch}}]{Nagle:1998}
\bibinfo{author}{\bibfnamefont{J.}~\bibnamefont{Nagle}},
  \bibinfo{author}{\bibfnamefont{H.}~\bibnamefont{Petrache}},
  \bibinfo{author}{\bibfnamefont{N.}~\bibnamefont{Gouliaev}},
  \bibinfo{author}{\bibfnamefont{S.}~\bibnamefont{Tristram-Nagle}},
  \bibinfo{author}{\bibfnamefont{Y.}~\bibnamefont{Liu}},
  \bibinfo{author}{\bibfnamefont{R.}~\bibnamefont{Suter}}, \bibnamefont{and}
  \bibinfo{author}{\bibfnamefont{K.}~\bibnamefont{Gawrisch}},
  \bibinfo{journal}{Phys. Rev. E} \textbf{\bibinfo{volume}{58}},
  \bibinfo{pages}{7769} (\bibinfo{year}{1998}).

\bibitem[{\citenamefont{Zhang et~al.}(1995)\citenamefont{Zhang, Sun,
  Tristram-Nagle, Headrick, Suter, and Nagle}}]{Zhang:1995}
\bibinfo{author}{\bibfnamefont{R.}~\bibnamefont{Zhang}},
  \bibinfo{author}{\bibfnamefont{W.}~\bibnamefont{Sun}},
  \bibinfo{author}{\bibfnamefont{S.}~\bibnamefont{Tristram-Nagle}},
  \bibinfo{author}{\bibfnamefont{R.~L.} \bibnamefont{Headrick}},
  \bibinfo{author}{\bibfnamefont{R.~M.} \bibnamefont{Suter}}, \bibnamefont{and}
  \bibinfo{author}{\bibfnamefont{J.~F.} \bibnamefont{Nagle}},
  \bibinfo{journal}{Phys. Rev. Lett.} \textbf{\bibinfo{volume}{74}},
  \bibinfo{pages}{2832} (\bibinfo{year}{1995}).

\bibitem[{\citenamefont{Chen et~al.}(1997)\citenamefont{Chen, Hung, and
  Huang}}]{Chen:1997}
\bibinfo{author}{\bibfnamefont{F.}~\bibnamefont{Chen}},
  \bibinfo{author}{\bibfnamefont{W.}~\bibnamefont{Hung}}, \bibnamefont{and}
  \bibinfo{author}{\bibfnamefont{H.}~\bibnamefont{Huang}},
  \bibinfo{journal}{Phys. Rev. Lett.} \textbf{\bibinfo{volume}{79}},
  \bibinfo{pages}{4026} (\bibinfo{year}{1997}).

\bibitem[{\citenamefont{Mason et~al.}(2001)\citenamefont{Mason, Nagle, Epand,
  and Katsaras}}]{Mason:2001}
\bibinfo{author}{\bibfnamefont{P.}~\bibnamefont{Mason}},
  \bibinfo{author}{\bibfnamefont{J.}~\bibnamefont{Nagle}},
  \bibinfo{author}{\bibfnamefont{R.}~\bibnamefont{Epand}}, \bibnamefont{and}
  \bibinfo{author}{\bibfnamefont{J.}~\bibnamefont{Katsaras}},
  \bibinfo{journal}{Phys. Rev. E} \textbf{\bibinfo{volume}{63}},
  \bibinfo{pages}{030902(R)} (\bibinfo{year}{2001}).

\bibitem[{\citenamefont{Pabst et~al.}(2003)\citenamefont{Pabst, Katsaras,
  Raghunathan, and Rappolt}}]{Pabst:2003}
\bibinfo{author}{\bibfnamefont{G.}~\bibnamefont{Pabst}},
  \bibinfo{author}{\bibfnamefont{J.}~\bibnamefont{Katsaras}},
  \bibinfo{author}{\bibfnamefont{V.~A.} \bibnamefont{Raghunathan}},
  \bibnamefont{and} \bibinfo{author}{\bibfnamefont{M.}~\bibnamefont{Rappolt}},
  \bibinfo{journal}{Langmuir} \textbf{\bibinfo{volume}{19}},
  \bibinfo{pages}{1716} (\bibinfo{year}{2003}).

\bibitem[{\citenamefont{Weast and Astle}(1979)}]{Weast:1979}
\bibinfo{editor}{\bibfnamefont{R.}~\bibnamefont{Weast}} \bibnamefont{and}
  \bibinfo{editor}{\bibfnamefont{M.}~\bibnamefont{Astle}}, eds.,
  \emph{\bibinfo{title}{CRC Handbook of Chemistry and Physics}}
  (\bibinfo{publisher}{CRC Press}, \bibinfo{address}{Boca Raton, FL},
  \bibinfo{year}{1979}), \bibinfo{edition}{60th} ed.

\end{thebibliography}

\end{document}